\documentclass[journal=jacsat,manuscript=article]{achemso}

\usepackage{chemformula} 
\usepackage[T1]{fontenc} 
\usepackage{soul} 
\usepackage{color}
\definecolor{vs}{rgb}{0.1,0.4,0.1}                  

\usepackage{hyperref}

\author{Michael I. Tribelsky}
\email{mitribel_at_gmail_dot_com}
\affiliation[M. V. Lomonosov Moscow State University]
{Faculty of Physics, M. V. Lomonosov Moscow State University, 119991 Moscow, Russia}
\author{Boris Y. Rubinstein}
\email{bru_at_stowers_dot_org}
\affiliation[Stowers Institute]
{Stowers Institute for Medical Research, 1000 E. 50th St., Kansas City, MO 64110, USA}

\title[The Poynting vector singularities]
  {The Poynting vector field generic singularities in resonant scattering of plane linearly polarized electromagnetic waves by subwavelength particles
  }

\abbreviations{2D,3D,IR,UV}

\begin{document}

%
%
%
%
%

\begin{abstract}
We present the results of a study of the Poynting vector field generic singularities at the resonant light scattering of a plane monochromatic linearly polarized electromagnetic wave by a subwavelength particle. We reveal the impact of the problem symmetry, the spatial dimension, and the energy conservation law on the properties of the singularities. We show that, in the cases when the problem symmetry results in the existence of an invariant plane for the Poynting vector field lines, a formation of a standing wave in the immediate vicinity of a singularity gives rise to a saddle-type singular point. All other types of singularities are associated with vanishing at the singular points, either (i) magnetic field, for the polarization plane parallel to the invariant plane, or (ii) electric field, at the perpendicular orientation of the polarization plane. We also show that in the case of two-dimensional problems (scattering by a cylinder), the energy conservation law restricts the types of possible singularities only to saddles and centers in the non-dissipative media and to saddles, foci, and nodes in dissipative. Finally, we show that dissipation affects the (i)-type singularities much stronger than the (ii)-type. The same conclusions are valid for the imaginary part of the Poynting vector in problems where the latter is regarded as a complex quantity. The singular points associated with the formation of standing waves are different for real and imaginary parts of this complex vector field, while all other singularities are common. We illustrate the general discussion by analyzing singularities at light scattering by a subwavelength Germanium cylinder with the actual dispersion of its refractive index.
%
%
\end{abstract}

Studies of light field singularities attract a great deal of attention and nowadays constitute a separate discipline called singular optics. Numerous  research papers, reviews, book chapters, and monographs present results in this subfield of optics; see, for example, Refs.~\cite{Angelsky2021,Yue2019,Gao2014,Dennis2009,Novitsky2009,Mokhun2007,luk2022}. In addition to the purely academic interest, the structure of the electromagnetic field in the immediate vicinity of a nanoparticle irradiated by a laser beam is of utmost importance for various problems of high-resolution spectroscopy, nanomaterial science, nanotechnologies, etc.; see, e.g., \cite{mueller2021surface,Xomalis2020,AlZoubi2022,Li2021} and references therein.
In its turn, singular points, their type, and position determine the topological structure of this field. On the other hand, it is known that the topological structure of the Poynting vector field in resonant light scattering by nanoparticles may be rather complicated. Specifically, the near-field wave zone and the particle itself may include singular points of different types~\cite{Luk2004:PRB,Zheludev2005:OE,OptJ_2006,Tribelsky:2006jf}. A detailed Poincar\'{e}-type classification of the Poynting vector field singular points was made by Novitsky and Barkovsky~\cite{Novitsky2009}. Authors of further publications basically are focused on the study of singularities in more complicated cases, such as, for example, in a non-diffractive tractor beam~\cite{Gao2014}, Bessel beam~\cite{Klimov2021}, etc.

Meanwhile, some fundamental questions related to the restrictions imposed by the space dimension, symmetry, and the energy conservation law on the type and properties of the singular points 
remain open. {To the best of our knowledge,} publication~\cite{Tribelsky2022} was one of the first attempts to answer these questions in the case of non-dissipative media. However, it is known that even weak dissipation may affect the topological structure of the Poynting vector field dramatically~\cite{tribelsky2022resonant,OptJ_2006,Luk2004:PRB}. Bearing in mind that a non-dissipative medium is an unachievable idealization never realized in actual physical systems, investigation of dissipative effects in this problem raises it to a much higher level. Here we present the results of this investigation.

\section{Results and discussion}

\subsection{The problem formulation}

The features of singular points discussed below are generic and valid for any shape of the scatterer and its optical properties. To \emph{illustrate} the obtained general results, we also consider several exactly solvable problems of the light scattering by a spatially uniform sphere and right circular cylinder~\cite{Bohren::1998}. The scattering particles are characterized by the complex permittivity $\varepsilon=\varepsilon'+i\varepsilon''$ and permeability $\mu = 1$, which is typical for the optical frequencies. We also assume that the scattering is passive, so that $\varepsilon''>0$. Active particles with population inversion and $\varepsilon''<0$ are not discussed, though the consideration can be readily generalized to this case by formal change of the sign of $\varepsilon''$.

We describe the electric and magnetic field by the expressions $\mathbf{E}(\mathbf{r})\exp(-i\omega t)$ and \linebreak $\mathbf{H}(\mathbf{r})\exp(-i\omega t)$, respectively so that the Poynting vector averaged over the field oscillations is
\begin{equation}\label{eq:Poynting_EH}
  \mathbf{S}=\frac{c}{16\pi}(\mathbf{E}^*\!\times\mathbf{H} +\mathbf{E}\times\mathbf{H}^*),
\end{equation}
where the asterisk stands for the complex conjugation.

At singular points, the direction of the Poynting vector is not determined, which can be the case only if $\mathbf{S}$ vanishes there. Then, in a generic case, according to Eq.~\eqref{eq:Poynting_EH}, at a singularity, either $\mathbf{E}=0$ (\emph{E-field-induced singularity}) or $\mathbf{H}=0$ (\emph{H-field-induced singularity}), or the r.h.s. of Eq.~\eqref{eq:Poynting_EH} vanishes, while neither $\mathbf{E}$, nor $\mathbf{H}$ do. (\emph{polarization-induced singularity})~\cite{Novitsky2009}. The latter means that the vector product $\mathbf{E}^*\!\times\mathbf{H}$ is purely imaginary, i.e., a standing wave is formed~\cite{Tribelsky2022}.

Regarding the Poynting vector field lines ({\it streamlines}), it is convenient to represent them in a parametric form: \mbox{$\mathbf{r} = \mathbf{r}(t)$,} where the ``time'' $t$ is not the actual time. It is just a dimensionless parameter. We introduce the dimensionless variables, normalizing $\mathbf{E}$, $\mathbf{H}$, and $\mathbf{S}$ on the corresponding values for the incident wave, while the dimensionless coordinate $\mathbf{r}_{\rm new} = \mathbf{r}_{\rm old}/R$, where $R$ is a certain characteristic spatial scale of the problem, whose specific choice may depend on the problem formulation; see below. Since in what follows, we use only dimensionless quantities, the subscripts ``old'' and ``new'' will be dropped.

By the definition of a field line for a vector field, the vector is tangential to the latter at any its point. It means that the ``velocity'' $d\mathbf{r}/dt$ is parallel to $\mathbf{S}$. The proportionality coefficient always may be turned to unity by the proper re-scaling of $t$. Then, the streamlines are defined by the following equation:
\begin{equation}\label{eq:stream_vector}
  \frac{d\mathbf{r}}{dt} = \mathbf{S}(\mathbf{r}).
\end{equation}

Equation \eqref{eq:stream_vector} must be supplemented by the energy conservation law: $-{\rm div}\,\mathbf{S}=q$, where $q$ is the power dissipated in the unite of volume~\cite{landau2013electrodynamics}. The conventional problem formulation corresponds to the incident wave coming from infinity and the scattered radiation going to infinity too. It is possible only if the scattering particle is embedded in a non-dissipative medium (in our case, it is a vacuum). Then, $q=0$ outside the scatterer. We discuss this case in our previous publication~\cite{Tribelsky2022}. Here we are interested in singular points situated within the scatterer too. For these points $q\neq 0$. Then, bearing in mind that for the problem in question $\mu=1$, in the selected dimensionless variables $q=\varepsilon''kR|\mathbf{E}|^2$, where $k=\omega/c$ is the incident wave wavenumber, and $c$ stands the speed of light in a vacuum\cite{landau2013electrodynamics}. In this case, the energy conservation law reads:
\begin{equation}\label{eq:divS}
  {\rm div}\,\mathbf{S}=-\varepsilon''kR|\mathbf{E}|^2.
\end{equation}

Equation~\eqref{eq:divS} imposes certain constraints on the roots of the singularity characteristic equation. The effects of the constraints increase with a decrease in the spatial dimension of the field pattern, i.e., with a reduction of the number of independent components of the Poynting vector, see below. Importantly, the r.h.s. of Eq.~\ref{eq:divS} does not depend on $\mathbf{H}$. Then, for the E-field-induced singularities, the r.h.s. of Eq.~\eqref{eq:divS} vanishes. At the same time, for other types of singularities, it remains finite, which means that dissipation affects the E-field-induced singularities much weaker than it does for H-field- and polarization-induced ones.

At the end of this subsection note that sometime it is convenient to introduce the complex Poynting vector $\hat{\mathbf{S}} = \frac{c}{16\pi} \mathbf{E}^*\!\times\mathbf{H}$, whose imaginary part characterizes the alternating flow of the so-called ``stored energy''~\cite{Jackson1998}. This imaginary part plays an important role in some problems of light-matter interaction \cite{bliokh2014magnetoelectric,Bliokh2014,Bekshaev2015,xu2019azimuthal,Khonina2021,Tang2010,Lininger2022}. Therefore, the field of the imaginary part of the complex Poynting vector is of interest too. Regarding singular points of this field, we can say that since, at the field-induced singularities of the conventional real Poynting vector, $\mathbf{S} \equiv {\rm Re}\, \hat{\mathbf{S}}$, the entire complex amplitude of either electric or magnetic field vanishes, at these points ${\rm Re}\, \hat{\mathbf{S}}={\rm Im}\, \hat{\mathbf{S}}=0$. That is to say, the singularities of ${\rm Re}\, \hat{\mathbf{S}}$ simultaneously are the ones for ${\rm Im}\, \hat{\mathbf{S}}$. However, this is not the case for polarization-induced singularities of ${\rm Re}\, \hat{\mathbf{S}}$. As it has been mentioned above, at these points, $\hat{\mathbf{S}}$ becomes a purely imaginary quantity, which, generally speaking, is not equal to zero. In other words, in the generic cases, the polarization-induced singular points for the vector field ${\rm Re}\, \hat{\mathbf{S}}$ remain regular points for ${\rm Im}\,\hat{\mathbf{S}}$ and vice versa. The presented analysis of singularities of ${\rm Re\,}\hat{\mathbf{S}}$ is readily transferred to the case of the field of ${\rm Im\,}\hat{\mathbf{S}}$ if required.

\subsection{Sphere}

To be more specific, we discuss here in detail several examples of singularities. First, we consider the scattering by a sphere. In this case, the field pattern is three-dimensional (3D). However, the symmetry of the problem dictates the existence of invariant planes. They are the planes passing through the center of the sphere  and either parallel to the polarization one (the plane of the vector $\mathbf{E}$ oscillations in the incident wave)~\cite{OptJ_2006,Tribelsky:2006jf,Luk2004:PRB,tribelsky2022resonant,Tribelsky2022,Zheludev2005:OE} or perpendicular to it. We select the conventional orientation of the coordinate frame~\cite{Bohren::1998} whose center coincides with the one for the sphere, plane $xz$ is parallel to the polarization plane, and the incident radiation propagates along the positive direction of the $z$-axis.

In the case of a subwavelength sphere, all singular points detected in this problem lie in the invariant plane parallel to the polarization plane~\cite{luk2007peculiaritiesCOLA,Tribelsky:2006jf,OptJ_2006,Luk2004:PRB,tribelsky2022resonant}. Presumably, this is related to the well-known fact that for such a scatterer, resonances may be associated with excitations of the so-called electric modes solely. All magnetic modes are off-resonant~\cite{Bohren::1998,born2013principles,Trib_JETP_1984}.

Let us consider the proximity of a singular point from this set. Importantly, while the pattern in the invariant plane is two-dimensional (2D), the problem itself is 3D. It means that though $S_y =0$ anywhere in this plane, $\partial S_y/\partial y \neq 0$ there. A generic singular point corresponds to a simple zero of $\mathbf{S}(\mathbf{r})$. Then, shifting the origin of the coordinate frame to the singularity, expanding $\mathbf{S}(\mathbf{r})$ about this point in the Tailor series, and keeping only the first non-vanishing terms, we obtain the following equation governing the streamlines:
\begin{eqnarray}
  \frac{dx}{dt} &=& S_x(x,y,z) \approx s_x^{(x)}x + s_{x}^{(y)}y + s_{x}^{(z)}z, \label{eq:dx/dt} \\
  \frac{dy}{dt} &=& S_y(x,y,z) \approx s_{y}^{(y)}y, \label{eq:dy/dt}\\
  \frac{dz}{dt} &=& S_z(x,y,z) \approx s_{z}^{(x)}x + s_{z}^{(y)}y + s_{z}^{(z)}z, \label{eq:dz/dt}
\end{eqnarray}
where $s_{x_n}^{(x_m)} \equiv \left(\frac{\partial S_{x_n}}{\partial x_m}\right)_{\! sin}$. Here the subscript {\it sin} means that the derivatives are taken at the singular point, and $x_m$ stand for any of the three components of vector {\bf r}.

To obtain the characteristic equation for system Eqs.~\eqref{eq:dx/dt}--\eqref{eq:dz/dt} we have to look for its solution in the form $x_n=x_{n0}\exp(\kappa t)$, $x_{n0}=const_n$. It gives rise to a cubic equation whose roots are as follows:
\begin{eqnarray}
  \kappa_1 &=& s_{y}^{(y)},\;\; \kappa_{2,3}=\gamma \pm \alpha, \label{eq:kappa_123} \\
  \gamma &=& \frac{s_{x}^{(x)}+s_{z}^{(z)}}{2},\;\; \alpha = \frac{\sqrt{\left(s_x^{(x)}-s_z^{(z)}\right)^2+4s_x^{(z)}s_z^{(x)}}}{2}. \label{eq:alpha_gamma_3D}
\end{eqnarray}
Now we recall, that $-{\rm div}\,\mathbf{S}=q$. In the discussed approximation ${\rm div}\,\mathbf{S} \approx s_{x}^{(x)} + s_{y}^{(y)} + s_{z}^{(z)}$. What is about $q$?

In the case of the H-field-induced or polarization-induced singularities $\mathbf{E}$ does not vanish at the singular point. Then, in the leading approximation $q = const$, i.e., $s_{x}^{(x)} + s_{y}^{(y)} + s_{z}^{(z)}=const$; see Eq.~\eqref{eq:divS}.

In the case of the E-field-induced singularities, $\mathbf{E}=0$ at the singular points. Then, for a generic singularity in its vicinity, $\mathbf{E}$ and $|\mathbf{E}|^2$ should be linear and quadratic forms of $x_n$, respectively. Since both sides of Eq.~\eqref{eq:divS} must have the same order of smallness in $x_n$, it means that the quadratic terms on the r.h.s. must be dropped, and in the leading approximation, $q=0$. That is to say, at the immediate vicinity of the E-field-induced singularity dissipation does not affect the streamline pattern, and the relation between the coefficients $s_{x_n}^{(x_m)}$ is the same as that for a non-dissipative medium, namely $s_{x}^{(x)} + s_{y}^{(y)} + s_{z}^{(z)}=0$.

Anyway, in all cases, we have only a single condition imposed on the seven non-zero entries of matrix $s_{x_n}^{(x_m)}$ (note, that, generally speaking, $s_{x_n}^{(x_m)} \neq s_{x_m}^{(x_n)}$ at $n \neq m$); \mbox{see Eqs.~\eqref{eq:dx/dt}--\eqref{eq:dz/dt}.} Plenty of parameters still remain free. As a result, the energy conservation law, Eq.~\eqref{eq:divS} does not make any type of singularities forbidden.

At the end of this subsection, it is relevant to derive certain universal relations, employing the integral form of the energy conservation law. For this purpose, employing the same local coordinate frame as that in Eqs.~\eqref{eq:dx/dt}--\eqref{eq:dz/dt},  we embed a singularity in a right circular cylinder with a small radius of the base $r$ and a small height $2y$, putting the singular point in the middle of the height of the cylinder. Bearing in mind that the total flux of the Poynting vector through the cylinder must be equal to the power dissipated in its volume, we obtain $2\pi r 2y \langle S_r \rangle + 2 \pi r^2 s_{y}^{(y)} y= - \pi r^2 2y q$. Here $\langle S_r \rangle$ is the radial component of the Poynting vector averaged over the azimuthal angle $\varphi$, and we take into account that the flux is calculated with respect to the outer normal and equals the power {\it leaking} from the cylinder. This expression may be rewritten as follows:
\begin{equation}\label{eq:<Sr>}
  \langle S_r\rangle = - \frac{rs_y^{(y)}+q}{2}.
\end{equation}
The above relation is valid in the vicinity of any singularity, regardless the specific type of the latter.
\subsection{Cylinder}
\subsubsection{General consideration}

\begin{figure}[ht]
\includegraphics[width=10.5 cm]{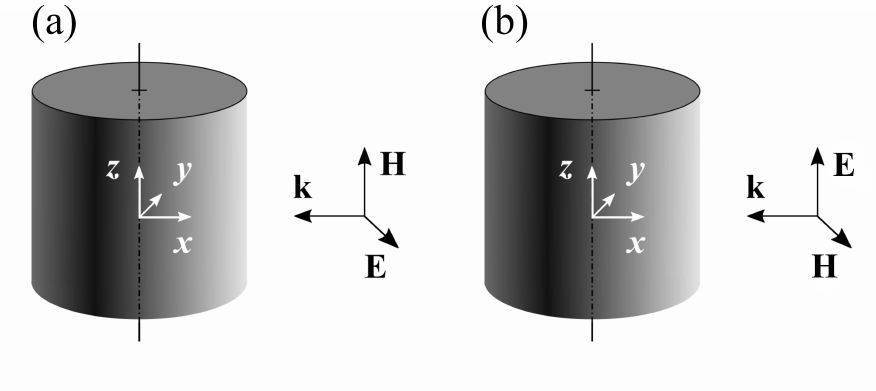}
\centering
\caption{{Mutual orientations of the cylinder and {\bf k}, {\bf E}, {\bf H} vectors of the incident wave. \mbox{TE polarization (a),} TM polarization (b).}\label{fig:TE_TM}}
\end{figure}

In the case of the scattering by an infinite cylinder, the problem is symmetric against an arbitrary translation along the axis of the scatterer, which effectively makes the problem 2D. Let us discuss here how the space dimension affects singularities. For the sake of simplicity, we restrict the consideration to the normal incidence of a purely TE- (vector $\mathbf{E}$ is perpendicular to the axis of the cylinder) or TM- (vector $\mathbf{H}$ is perpendicular to the axis of the cylinder) polarized wave and right circular homogeneous cylinder. According to the conventional notations, the axis of the cylinder is selected as the $z$-axis of the cylindrical coordinate frame, and the wave vector of the incident wave, $\mathbf{k}$ is supposed to be antiparallel to the $x$-axis~\cite{Bohren::1998}; see Fig.~\ref{fig:TE_TM}

In both polarizations, the problem symmetry makes the $xy$-plane invariant for the streamline pattern. The streamlines in this plane are described by a 2D version of Eq.~\eqref{eq:stream_vector}. In the local coordinate frame, whose origin coincides with the singularity in question, in the vicinity of the singularity, Eq.~\eqref{eq:stream_vector} transforms into the following:
\begin{eqnarray}
  \frac{dx}{dt} &=& S_x(x,y) \approx s_x^{(x)}x + s_{x}^{(y)}y, \label{eq:2D_dx/dt} \\
  \frac{dy}{dt} &=& S_y(x,y) \approx s_{y}^{(x)}x + s_{y}^{(y)}y, \label{eq:2D_dy/dt}
\end{eqnarray}
while the application of the energy conservation law gives rise to the constraint
\begin{equation}\label{eq:2D_divS=0}
  s_x^{(x)}+s_y^{(y)}=-q.
\end{equation}

Employing Eq.~\eqref{eq:2D_divS=0}, the roots of the characteristic equation once again may be written as $\kappa_{1,2} = \gamma \pm \alpha$. However, now
\begin{equation}\label{kappa12_cyl}
  \gamma = -\frac{q}{2},\;\; \alpha = \frac{\sqrt{\left(2s_x^{(x)}+q\right)^2+4s_x^{(y)}s_y^{(x)}}}{2}
\end{equation}
It imposes certain restrictions on the possible types of singularities; see Table~\ref{tbl:Singularities_cyl}. The table indicates that the strictest reductions of the variety of singular points occur in the non-dissipative limit ($q=0$), when only saddles and centers can come into being. This conclusion agrees with the results of various computer simulations of the streamline pattern for a cylinder; see, e.g., Refs.\cite{luk2007peculiaritiesCOLA,OptJ_2006}.

\begin{table}[ht]
  \caption{Singularities}
  \label{tbl:Singularities_cyl}
\begin{tabular}{lll}
\hline
\multicolumn{2}{l}{Conditions} & Type of singularity \\
\hline
$\left(2s_x^{(x)}+q\right)^2+4s_x^{(y)}s_y^{(x)}<0$; & $q=0$ & Center  \\
$\left(2s_x^{(x)}+q\right)^2+4s_x^{(y)}s_y^{(x)}>0$; & $q=0$ & Saddle \\
$\left(2s_x^{(x)}+q\right)^2+4s_x^{(y)}s_y^{(x)}<0$; & $q>0$ &  Focus \\
$\left(2s_x^{(x)}+q\right)^2+4s_x^{(y)}s_y^{(x)}>0$; & $0<q<2\alpha$ & Saddle \\
$\left(2s_x^{(x)}+q\right)^2+4s_x^{(y)}s_y^{(x)}>0$; & $q>2\alpha$ & Node
\end{tabular}
\end{table}

\subsubsection{Examples}

Here we present specific examples illustrating the above general consideration. To be close to an actual experimental case, we consider the light scattering by a Germanium right circular cylinder with the actual dispersion of the refractive index $n=n'(\lambda)+in''(\lambda)$ for this material~\cite{Polyanskiy}, where $\lambda$ stands for the incident wavelength in a vacuum. The choice of Germanium is convenient because of its high values of $n'$ and strong dispersion of $n''$ in the visible and near IR ranges of the spectrum. Regarding the choice of the characteristic spatial scale of the problem $R$, to make a comparison of field patterns for different values of the wavelength and radius of the cylinder, it is convenient to select for $R$ the radius of the base of the cylinder so that the dimensionless radius is always equal to unity.

\begin{figure}[!htb]
  \centering
  \includegraphics[width=\textwidth]{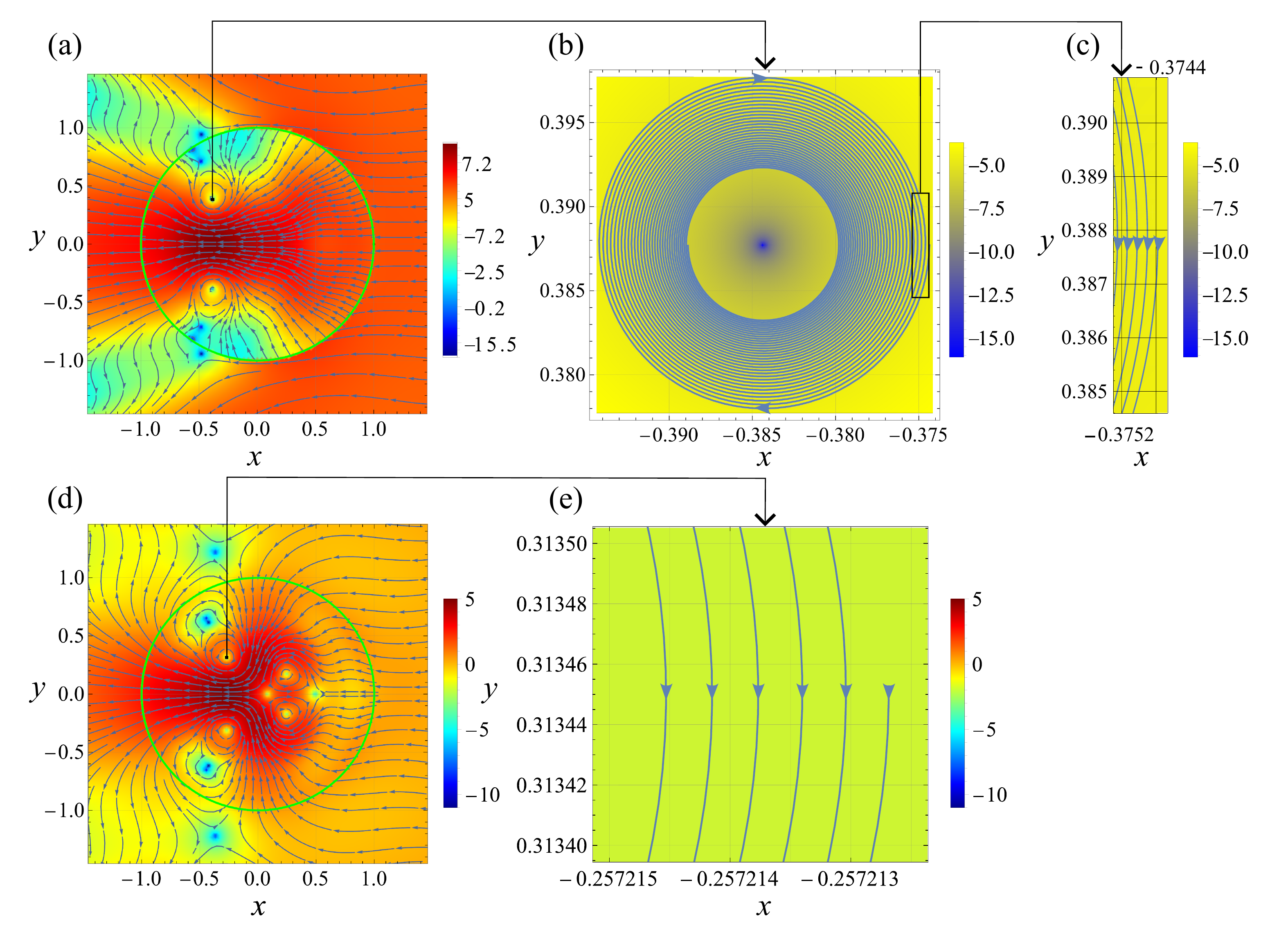}
  \caption{Light scattering by a Germanium cylinder in a vacuum. $\lambda=\lambda_1=1590$ nm. \mbox{$\varepsilon(\lambda_1)\approx 17.775+i0.024$}. Streamlines for the Poynting vector field. The pattern is symmetric with respect to the line $y=0$. The green circles in (a) and (d) designate the surface of the cylinder. The color indicates the value of $\ln|\mathbf{S}|^2$; see the color bars. The panels in the upper and low rows correspond to the TE and TM polarization of the incident wave, respectively. Panel (b) is a zoom of the vicinity of the singular point shown in (a) as a small black rectangle. The same is true for (d) and (e). Panel (c) is a zoom of the region marked in (b) as a rectangle. See text for details.
  }\label{fig:1590}
\end{figure}
Specifically, we select the two values of $\lambda$: $\lambda_1 = 1590$~nm and $\lambda_2 = 1494$~nm. At these wavelengths Germanium has the following values of permittivity~\cite{Polyanskiy}: \mbox{$\varepsilon(\lambda_1)\approx 17.775+i0.024$} and  \mbox{$\varepsilon(\lambda_2)\approx 17.983+i0.483$.} Regarding $R$, we select it so that the size parameter $kR$ keeps the same value 1.62 at both values of $\lambda$. This value of the size parameter lies in the vicinity of the dipolar resonances at both (TE and TM) independent polarizations of the incident wave.

Since \mbox{$\varepsilon'(\lambda_1)\approx \varepsilon'(\lambda_2)$}, while $\varepsilon''(\lambda_2)$ is more than twenty times larger than $\varepsilon''(\lambda_1)$, such a choice makes it possible to study the effects of dissipation solely and compare the results for TE and TM polarizations at almost fixed values of the other problem parameters. Note also that for the given value of the size parameter $R/\lambda \approx 0.26$, i.e., the cylinder is a subwavelength scatterer.
\begin{figure}[!htb]
  \centering
  \includegraphics[width=\textwidth]{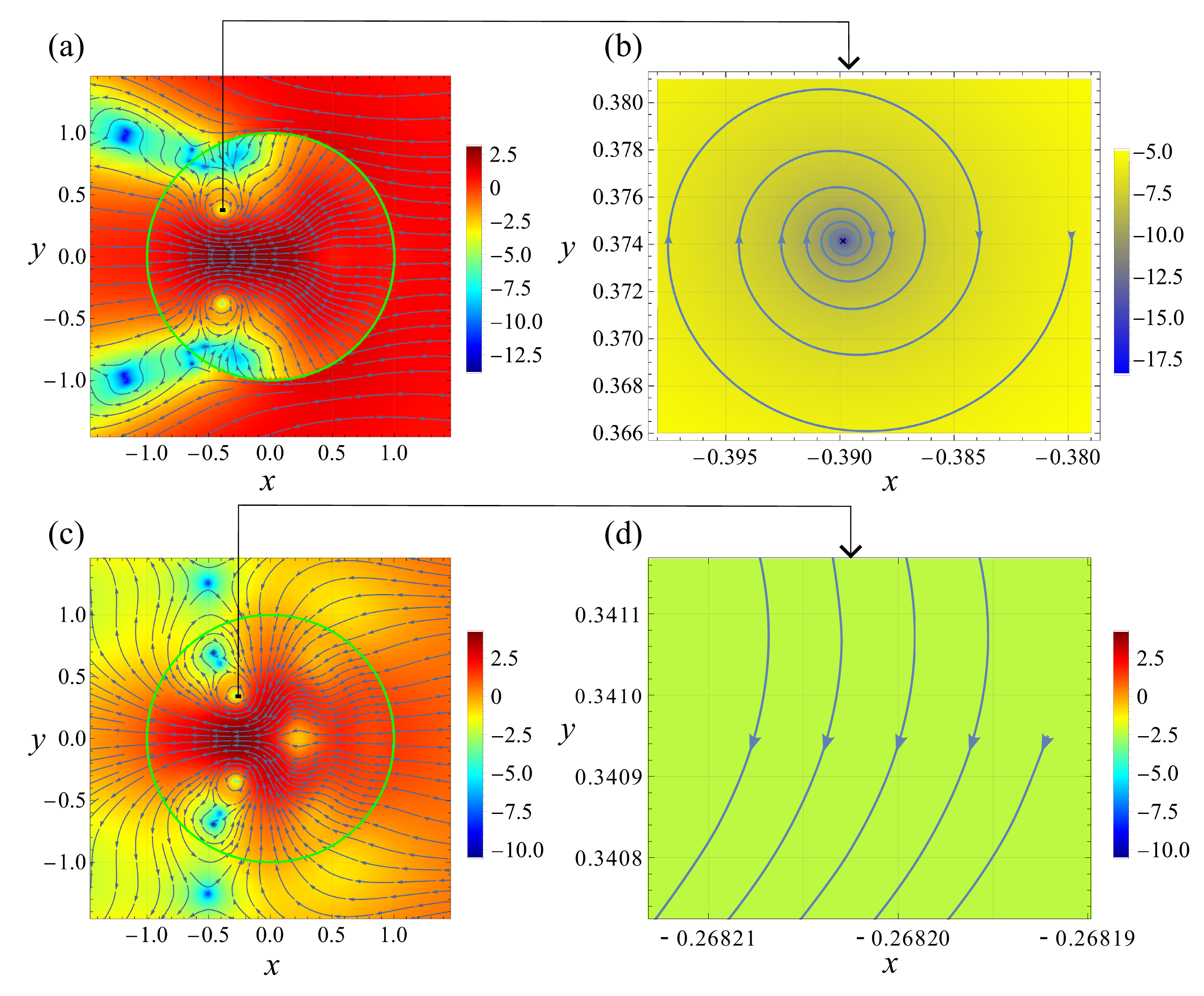}
  \caption{The same as that in Fig.~\ref{fig:1590} but at $\lambda=\lambda_2=1494$ nm; \mbox{$\varepsilon(\lambda_2)\approx 17.983+i0.483$.}}\label{fig:1494}
\end{figure}

Figures \ref{fig:1590}, \ref{fig:1494} show the visualization of the analytical solution to the scattering problem at the indicated values of the problem parameters. Specifically, \mbox{Fig.~\ref{fig:1590}(a), (d)} and \mbox{Fig.~\ref{fig:1494}(a), (c)} show the general structure of the Poynting vector field for the TE and TM polarizations of the incident wave, respectively.

Due to the same value of the size parameter, relatively low dissipation, close values of $\lambda_{1,2}$ and the selection of the problem parameters so that in all cases, $\lambda_{1,2}$ lie in the vicinity of the corresponding dipolar resonance, the patterns in \mbox{Fig.~\ref{fig:1590}(a), (d)} and \mbox{Fig.~\ref{fig:1494}(a), (c)} exhibit remarkable similarity. In particular, there are several singular points both inside and outside the cylinder situated similarly. In agreement with the above general discussion, $\mathbf{S}=0$ at each of them.

In the specific cases inspected in our previous publication~\cite{Tribelsky2022} it was shown that, in the case of a cylinder, {the polarization-induced singularities are saddles, while all other types of singular points are field-induced. Moreover, if the plane of the oscillations of vector $\mathbf{E}$ in the incident wave is parallel to the cylinder base (and hence the plane of vector $\mathbf{H}$ oscillations is perpendicular to it), the field-induced singularities occur owing to the vanishing of the magnetic field at the singular point. Vice versa: at the perpendicular orientation of the polarization plane, when the plane of the oscillations of vector $\mathbf{H}$  is parallel to the base of the cylinder,} the singularities are E-field-induced. If this feature is generic, it allows to change the types of the field-induced singularities from H to E just by changing the incident wave polarization from TE to TM. Our calculations show that this is the case indeed. Thus, the field-induced singularities in Figs.~\ref{fig:1590}(a) and \ref{fig:1494}(a) are of the H-type, while those in Figs.~\ref{fig:1590}(d) and \ref{fig:1494}(c) are of the E-type. The saddles are polarization-induced: none of vectors $\mathbf{E}$ and  $\mathbf{H}$ vanishes at the saddles. Importantly, the region of the standing wave formation in the vicinity of the saddles is substantially subwavelength relative to the incident wave.

Now note that despite the apparent similarity, there is a drastic difference between Figs.~\ref{fig:1590} and \ref{fig:1494}. To demonstrate that we perform a detailed inspection of the singularities marked in Fig.~\ref{fig:1590}(a), (d) and Fig.~\ref{fig:1494}(a), (c) with the small black rectangles. The coordinates of these singular points are presented in Table~\ref{tbl:Coordinates}, where ellipses denote dropped decimals.
\begin{table}[ht]
  \caption{Coordinates of singularities}
  \label{tbl:Coordinates}
\begin{tabular}{lcc}
\hline
Figure & $x_{\rm sin}$       & $y_{\rm sin}$   \\
\hline
Fig.~\ref{fig:1590}(a) & -0.384... & 0.387...   \\
Fig.~\ref{fig:1590}(d) & -0.267... & 0.313...  \\
Fig.~\ref{fig:1494}(a) & -0.389... & 0.374...   \\
Fig.~\ref{fig:1494}(c) & -0.278... & 0.340...
\end{tabular}
\end{table}

In Fig.~\ref{fig:1590}(a), (d) and Fig.~\ref{fig:1494}(a), (c) the singularities look like centers, but actually they are foci! To make sure of this, we zoom into the close vicinity of the singularities and inspect the behavior of streamlines situated there. Specifically, for each singularity we select the initial point, whose $y$-coordinates equals $y_{\rm sin}$ and $x$-coordinate is $x_{\rm sin} + d$, where in all cases $d=0.01$. Then we obtain the streamline originated in this initial point by numerical integration of Eq.~\eqref{eq:stream_vector}, where the $\mathbf{S}(\mathbf{r})$ is described by the well-known exact solution of the scattering problem~\cite{Bohren::1998}.

From now on, it is convenient to discuss each case separately. Fig.~\ref{fig:1590}(b), (c) indicates that the given streamline is a converging spiral with gradually decreasing pitch; see Eq.~\eqref{kappa12_cyl}. To be able to compare the cases quantitatively, it is relevant to give the initial value of the pitch, corresponding to the difference between the $x$ coordinate of the initial point and the one for the streamline after its first return to the same value of $y=y_{\rm sin}$. The corresponding value is $2.4208...\times 10^{-4}$.

In the case of the TM-polarized incident wave, the first pitch of the spiral is \mbox{$2.4875...\times 10^{-7}$}. It is so tiny that we cannot show the entire spiral with the proper resolution, as it is done in Fig.~\ref{fig:1590}(b). Therefore, only a mini-window accommodating fractions of the streamline, corresponding to a few of its first turns, is zoomed; see Fig.~\ref{fig:1590}(e).

The 20-fold increase in dissipation, occurring at $\lambda = 1494$ nm, affects the values of the first pitch accordingly; namely, we have it equal to $4.0456...\times 10^{-3}$ for the TE-polarized wave and $3.8679...\times 10^{-6}$ for TM-polarized; see Fig.~\ref{fig:1494}.
Thus, in the entire agreement with the above general discussion, the impact of dissipation on the structure of the field-induced singularities, in the case of the TE-polarized incident wave, is relatively strong. In contrast, for the TM polarization it is much weaker.

At the end of this section, it is relevant to make several comments related to the overall structure of the field pattern. First, note that while in the case of the TE-polarized incident wave, the streamlines undergo ``refraction'' at the surface of the cylinder, this is not the case for the TM polarization, cf. Figs.~\ref{fig:1590}(a), \ref{fig:1494}(a) and Figs.~\ref{fig:1590}(d), \ref{fig:1494}(c). To explain this peculiarity, note that the boundary conditions for the Maxwell equations stipulate continuity of the {\it tangential\/} to the surface components of $\mathbf{E}$ and $\mathbf{H}$. As for the normal components, they may undergo discontinuity~\cite{Bohren::1998}. However, just {\it may} not {\it must}! Regarding the electric field, at the surface of the cylinder, its normal component satisfies the condition $E_n^{\rm (out)}=\varepsilon E_n^{\rm (in)}$, where superscripts (in) and (out) designate the field outside and inside the cylinder, respectively. In other words, $E_n$ is discontinuous indeed.

However, since the permeability of the cylinder is unity, for magnetic field, the material of the cylinder is {\it indistinguishable\/} from a vacuum, and the normal component of magnetic field remains {\it continuous\/} at the surface. Then, for the TM polarization, when the normal component of the electric field of the incident wave identically equals zero, all components of $\mathbf{E}$ and $\mathbf{H}$, and hence the Poynting vector too, remain continuous at the surface of the cylinder. For the TE polarization vector $\mathbf{H}$ remains continuous, while vector $\mathbf{E}$ does not. Then, vector $\mathbf{S}$ undergoes the ``refraction''  due to the discontinuity of the normal component of $\mathbf{E}$.

The second comment is related to the general effects of dissipation. The conventional method to build the exact solution to the scattering problem is the multipolar expansion~\cite{Bohren::1998}. In this case, every partial field component outside the scatterer is presented as a sum of the incident and scattered partial waves, while the field inside the scatter is not split into an analogous sum. However, due to the linearity of the problem, we can always single out from the field inside the particle the components equal to the corresponding multipoles of the incident field, considering the rest as the field excited in the particle. The dissipation does not affect the former but does the latter. Namely, with an increase in dissipation, the amplitude of the incident wave remains fixed, while the amplitudes of the fields of the excited multipoles decrease both outside the particle and within it. On the other hand, the Poynting vector vanishes at singular points. It imposes a specific restriction on the amplitudes of the excited multipoles, which cannot be too small to be able to compensate the field of the incident wave.

From these arguments, it is clear that an increase in dissipation must eventually result in a decrease in the number of singular points both inside and outside the scatterer. Topologically it occurs owing to the merging of singularities resulting in their annihilation. Indeed, comparison of Fig.~\ref{fig:1590}(d) and Fig.~\ref{fig:1494}(c) reveals the vanishing in the latter of four singular points located in Fig.~\ref{fig:1590}(d) at $x>0$. Instead of them, only minor distortion of the streamlines remains.

\section{Conclusions}

Thus, in this study, we have inspected the effects of restrictions imposed on singularities of the Poynting vector field by the problem symmetry, energy conservation law, and spatial dimension. We have confirmed the generic nature of the preliminary results reported in our previous publication~\cite{Tribelsky2022}. Specifically, we have shown that {in 2D problems} {the polarization-induced singularities are saddles and} explained by the formation of standing waves in their immediate vicinity. In contrast, all other types of singularities are associated with the vanishing of either electric or magnetic fields at the singular points.

Notably, at light scattering by a subwavelength particle, the size of the standing wave formation area in proximity of saddles is {\it very much} smaller than the wavelength of the incident radiation. Regarding the other types of singularities associated with the vanishing of the fields $\mathbf{E}$ or $\mathbf{H}$, the vanishing affects only the field supplementary to the one lying in the invariant plane. In other words, if the polarization plane is parallel to the invariant plane, the vector $\mathbf{H}$ vanishes; if it is perpendicular to the invariant plane, vector $\mathbf{E}$ does.

We also show that the impact of the spatial dimension on the singularities increases with the transition from a three-dimensional scattering problem (sphere) to a two-dimensional (cylinder). While for the former, all types of singular points are possible, in the latter, the possible types are strictly limited, see Table~\ref{tbl:Singularities_cyl}.

We also show that while dissipation strongly affects the field lines in the vicinity of singularities related to the vanishing of the magnetic field, it makes a much weaker impact on the singularities associated with the vanishing of the electric field. This asymmetry between $\mathbf{E}$ and $\mathbf{H}$ is explained by the fact that the dissipated power vanishes at the E-field-induces singularities regardless of the value of dissipative constant, $\varepsilon''$; see expression \eqref{eq:divS}.  At the same time, at the H-field-induced singular points, the dissipation remains finite.

We illustrate the general {consideration} by detailed inspection of light scattering by a Germanium cylinder with actual dispersion of its permittivity~\cite{Polyanskiy}. All obtained generic conclusions excellently agree with the manifestation of the scattering in this specific case.

If we consider the Poynting vector as a complex quantity~\cite{Jackson1998,bliokh2014magnetoelectric,Bliokh2014,Bekshaev2015,xu2019azimuthal,Khonina2021,Tang2010,Lininger2022}, the same conclusions are valid for its imaginary part. Notably, while the field-induced singularities are common for both real and imaginary parts of this complex vector field, the polarization-induced singularities associated with the formation of standing waves for real and imaginary parts are different.

We hope these results shed new light on the fundamental problem of the nano-scale energy circulation at resonant light scattering by subwavelength particles and may be helpful in numerous applications and technologies.

\section{Methods}

The presented analysis is based on the exact solutions to the related problems;
see, e.g., Ref.~\cite{Bohren::1998}. The symbolic and numerical calculations, as well as the visualization of the results, are made with the help of the Wolfram {\it Mathematica} software.


\begin{acknowledgement}
MT acknowledges the financial support of the Russian Foundation for Basic Research (Projects No. 20-02-00086) for the analytical study and the Russian Science Foundation (Project No. 21-12-00151) for the computer calculations.
\end{acknowledgement}

%
%
%
%
\bibliography{Singularities_JCP}
\end{document}